\documentclass[traditabstract]{aa} 
\usepackage{subfigure}                                   
\usepackage{epsfig}
\usepackage{color}
\providecommand{\tabularnewline}{\\}
\usepackage{graphicx}
\usepackage{txfonts}
\begin{document}
   \title{Overabundance of $\alpha$-elements in exoplanet-hosting stars \thanks{Based on observations collected at 
the La Silla Parana Observatory,
ESO (Chile) with the HARPS spectrograph at the 3.6-m telescope (ESO
runs ID 72.C-0488, 082.C-0212, and 085.C-0063).}}

\author{V.~Zh.~Adibekyan\inst{1}\and N.~C.~Santos\inst{1,2}\and  S.~G.~Sousa\inst{1,3}\and G.~Israelian\inst{3,4}
\and E.~Delgado~Mena\inst{1} \and J.~I.~Gonz\'{a}lez Hern\'{a}ndez \inst{3,4} \and M.~Mayor \inst{5} \and C.~Lovis \inst{5} 
\and S.~Udry \inst{5}}

\institute{Centro de Astrof\'{\i}ísica da Universidade do Porto, Rua das Estrelas,
4150-762 Porto, Portugal\\
\email{Vardan.Adibekyan@astro.up.pt}\and Departamento de F\'{\i}ísica
e Astronomia, Faculdade de Ci\^{e}ncias da Universidade do Porto, Portugal\and Instituto
de Astrof\'{\i}sica de Canarias, 38200 La Laguna, Tenerife, Spain\and Departamento de Astrof{\'\i}sica, Universidad de La Laguna,
38206 La Laguna, Tenerife, Spain 
\and Observatoire de Gen\`{e}ve, Universit\'{e} de Gen\`{e}ve, 51 Ch. des Mailletes, 1290 Sauverny, Switzerland}

   \date{Received ...; Accepted...}

 
 \abstract
{We present the results for a chemical abundance analysis between planet-hosting and stars without planets for 12 refractory elements 
  for a total of 1111 nearby FGK dwarf stars observed within the context of the HARPS GTO programs. 
  Of these stars, 109 are known to harbour high-mass planetary companions and 26 stars are hosting exclusively Neptunians and super-Earths. 
  We found that the [X/Fe] ratios for Mg, Al, Si, Sc, and Ti both for giant
  and low-mass planet hosts are systematically higher than those of comparison stars  at low metallicities ([Fe/H] $\lesssim$ from -0.2 to 0.1 dex depending on the element). 
  The most evident discrepancy between planet-hosting and stars without planets is observed for Mg.
  Our data suggest that the planet incidence is greater among the thick disk population 
  than among the thin disk for mettallicities bellow -0.3 dex. After examining the [$\alpha$/Fe] trends of the planet host and non-host samples we conclude 
  that a certain chemical composition, and not the Galactic birth place of the stars, is the determinating factor for that.
  The inspection of the Galactic orbital parameters and kinematics of the planet-hosting stars shows that Neptunian hosts tend to belong to the ''thicker`` 
  disk compared to their high-mass planet-hosting counterparts. 
  We also found that Neptunian hosts follow the distribution of high-$\alpha$ stars in the \emph{UW} vs \emph{V} velocities space, but they are more 
  enhanced in Mg than high-$\alpha$ stars without planetary companions.
  Our results indicate that some metals other than iron may also have an important contribution to planet formation if the amount of iron is low. These results
  may provide strong constraints for the models of planet formation, especially for planets with low mass.}{}


\maketitle
%

\section{Introduction}

Since the first discovery of an exoplanet around a solar-like star, (Mayor \& Queloz \cite{Mayor-95}), more than 750 exoplanets
have been discovered. More than one hundred of these planets have masses less than 30 $\emph{M}_{\oplus}$. With the ever increasing number 
comes the realization that the nature of these planets is diverse, thereby adding complexities to an already complicated problem. 

There are essentially two main and competitive planet formation models: the core-accretion model (e.g. Pollack et al. \cite{Pollack-96}; 
Ida \& Lin \cite{Ida-04}; Mordasini et al. \cite{Mordasini-09}) and the disk instability model (e.g. Boss \cite{Boss-97};\cite{Boss-02}).
The tendency for planets to orbit metal-rich stars (Gonzalez \cite{Gonzalez-98}; Gonzalez et al. \cite{Gonzalez-01}; 
Santos et al. \cite{Santos-01,Santos-03,Santos-04}; Fischer \& Valenti \cite{Fischer-05}; Sousa et al. \cite{Sousa-08}; Neves et al.
\cite{Neves-09}; Johnson et al. \cite{Johnson-10}; Sousa et al. \cite{Sousa-11a}; Adibekyan et al. 2012 submitted) lends strong support to the first scenario, 
according to which planets grow through accretion of solid, metal-rich material to form massive cores. The discovery of several planets orbiting 
metal-poor stars (e.g. Cochran et al. \cite{Cochran-07}; Santos et al. \cite{Santos-10}) shows, however, that giant planet formation is not completely inhibited in 
the metal-poor regime (see also discussion in Santos et al. \cite{Santos-04,Santos-11}). 
Recent studies showed that the well-established metallicity  correlation seems to be only observed for giant planets. 
The stars hosting Neptunes and super-Earth-class planets show typical metallicities of the stars without planets (e.g. Udry et al. \cite{Udry-06};
Sousa et al. \cite{Sousa-08}; Ghezzi et al. \cite{Ghezzi-10}; Mayor et al. \cite{Mayor-11}; Sousa et al. \cite{Sousa-11a}). 
This gives us interesting hints about the planet formation processes.

In the conventional core-accretion scenario, the formation of planetesimals starts from the condensation of heavy elements (metals). 
In this context, the study of the heavy element abundances is very important. Previous studies that aimed to clarify whether the planet-hosting stars (PHS) 
are different from stars without planets in their content of individual heavy elements (other than iron) yielded clontradictory results. 
Although most studies showed no significant differences in the overall trends of {[}X/Fe{]} between PHSs and stars
without any known planetary-mass companions (e.g. Takeda \cite{Takeda-07}; Bond et al. \cite{Bond-08}; Neves et al. \cite{Neves-09}; Gonz\'{a}lez
Hern\'{a}ández et al. \cite{Gonzalez Hernandez-10}; Delgado Mena et al. \cite{Delgado-10}), some works have reported possible enrichment
in some species (Gonzalez et al. \cite{Gonzalez-01}; Santos et al. \cite{Santos-00}; Sadakane et al. \cite{Sadakane-02}; 
Bodaghee et al. \cite{Bodaghee-03}; Fischer\& Valenti \cite{Fischer-05}; Beir\~{a}o et al. \cite{Beirao-05}; Gilli et al. \cite{Gilli-06}; 
Bond et al. \cite{Bond-06}; Robinson et al. \cite{Robinson-06}; Gonzalez \& Laws \cite{Gonzalez-07};
Brugamyer et al. \cite{Brugamyer-11}; Kang et al. \cite{Kang-11}).

Although theoretical modeling suggests that metallicity is a key parameter
of planet formation, Haywood (\cite{Haywood-08}), studying the memberships
of PHSs to different stellar populations, proposed that the presence
of giant planets might be primarily a function of a parameter linked to 
galactocentric radius, but not metallicity and the apparent correlation 
between metallicity and the detection of planets is a natural consequence of that.
Haywood (\cite{Haywood-09}) suggested that this parameter, which depends on the distance from the Galactic Center,
could be the density of molecular hydrogen.
If the rate of giant planets does not depend on metallicity, then the core-accretion theory of planet formation 
(e.g. Ida \& Lin \cite{Ida-05}; Mordasini et al. \cite{Mordasini-12}) loses its most important observational support, and in turn strengthens the  gravitational 
instability theory (e.g. Boss \cite{Boss-01}; Cai et al. \cite{Cai-06}).
The kinematics of extra-solar planet hosts and their relation to different stellar populations and 
moving groups have been discussed  in some other works (e.g. Gonzalez \cite{Gonzalez-99}; Reid \cite{Reid-02};
Barbieri \& Gratton \cite{Barbieri-02}; Santos et al. \cite{Santos-03}; Ecuvillon et al. \cite{Ecuvillon-07}; Neves et al. \cite{Neves-09}; Gonzalez \cite{Gonzalez-09}).
The results do not allow one to reach any clear conclusion.

The study of extrasolar planets requires very high quality data. In particular, very high-precision radial-velocity
measurements are needed to detect planets. Likewise, finding possible abundance differences (sometimes
a very subtle task) between stars with and without planets also requires using large stellar samples with accurate and
homogeneous abundance determinations.
In this paper, we present a uniform spectroscopic analysis of 
1111 FGK dwarfs observed within the context of the HARPS GTO planet search program. Of these stars, 109 are known to harbour high-mass planetary 
companions and 26 stars are hosting exclusively Neptunians and super-Earths. The large size and homogeneity of this sample enables a very robust 
differential comparison of the chemical abundances of the stars  with and without planets with minimal internal uncertainties. 
The relatively high number of low-mass planet hosts in the sample allowed us to study their chemical properties separately from their Jupiter-mass 
planet hosting counterparts. The introduction of the sample and the methods of the chemical abundance determination and analysis are described
in details in our previous paper (Adibekyan et al. 2012, submitted). 
This paper is organized as follows: In Sect. 2 we briefly introduce the physical properties of the stars in the sample. 
The study of the abundances of refractory elements relative to iron in exoplanet-hosting stars can be found in Sect. 3. The kinematical
properties of the exoplanet-hosting stars is presented in Sect. 4. Finally, in Sect. 5  , we draw our
main conclusions.

\section{The sample}

The sample used in this work consists of 1111 FGK  stars observed with the HARPS spectrograph (Mayor et al. \cite{Mayor}) at the
ESO 3.6 m telescope (La Silla, Chile). The stars are slowly rotating, non-evolved, and in general have a low level of activity. 
Most stars have effective temperatures 4600 \emph{K} $\leq$ \emph{$T{}_{\mathrm{eff}}$} $\leq$ 6300 \emph{K} (only 4\% are outside of this region)
surface gravities 4 $\leq$ $\log\,g$ $\leq$ 5 dex (the number of ``outliers'' again is very small, only 25 stars with $\log\, g$ $<$ 4 dex)
and they lie in the metallicity range of -1.39 $\leq$ {[}Fe/H{]} $\leq$ 0.55 dex.

Precise stellar parameters for all stars were determined in the same manner and from the same spectra as used in our previous study. 
For details we refer the reader to Sousa et al. (\cite{Sousa-08,Sousa-11a,Sousa-11b}). The typical uncertainties in the atmospheric parameters are of 
about 30 \emph{K }for \emph{$T{}_{\mathrm{eff}}$} , 0.06 dex for $\log\, g$, and 0.03 dex for {[}Fe/H{]}.

Elemental abundances for 12 elements (Na, Mg, Al, Si, Ca, Ti, Cr,
Ni, Co, Sc, Mn, and V) were determined using a local thermodynamic equilibrium (LTE) 
analysis relative to the Sun with the 2010 revised version of the
spectral synthesis code MOOG%
\footnote{The source code of MOOG2010 can be downloaded at http://www.as.utexas.edu/\textasciitilde{}chris/moog.html%
} (Sneden \cite{Sneden}) and a grid of Kurucz ATLAS9 plane-parallel model atmospheres (Kurucz et al. \cite{Kurucz}).
The reference abundances used in the abundance analysis were taken from Anders \& Grevesse (\cite{Anders-89}).
The line list and atomic parameters of Neves et al. (\cite{Neves-09}) were used adding one CaI line and excluding
five NiI lines, two SiI lines, two TiII lines, and five TiI lines (Adibekyan et al. 2012, submitted). 
The equivalent widths were automatically measured with the ARES%
\footnote{The ARES code can be downloaded at http://www.astro.up.pt/sousasag/ares%
} code (Automatic Routine for line Equivalent widths in stellar Spectra
- Sousa et al. \cite{Sousa-07}). The final abundance for each star and element was calculated to be the average value of 
the abundances given by all lines detected in a given star and element. 
Individual lines for a given star and element with a line dispersion higher than a factor of two than the \emph{rms} were excluded.
The total uncertainties in the [X/H] abundances and [X/Fe] ratios do not exceed 0.04 and 0.03 dex for stars with 
\emph{$T{}_{\mathrm{eff}}$} = \emph{T$_{\odot}$$\pm$$500$ K}, respectively, and are less than 0.1 dex (except ScI, TiI and VI) for the stars
with temperatures very different from that of the Sun.

With our large stellar sample, we were able to detect and correct the [X/Fe] trends with \emph{$T{}_{\mathrm{eff}}$} 
for some elements. For AlI and CoI we observed a systematic trend with \emph{$T{}_{\mathrm{eff}}$} in all temperature ranges, and for
TiI, ScI, V, CrII, and Na we observed a trend with \emph{$T{}_{\mathrm{eff}}$} in the low-temperature domain.

As a check of our method and analysis, we compared our derived abundances with those obtained by  
Bensby et al. (\cite{Bensby-05}), Valenti \& Fisher (\cite{Valenti-05}), Gilli et al. (\cite{Gilli-06}), and Takeda (\cite{Takeda-07}) for stars in common with these works.
In genereal we found a good agreement with these previous studies.

For more details about the sample, abundance determination methods, analysis, and errors we refer the reader to Adibekyan et al. (2012, submitted).

\begin{figure*}
\centering
\includegraphics[angle=270,width=1\linewidth]{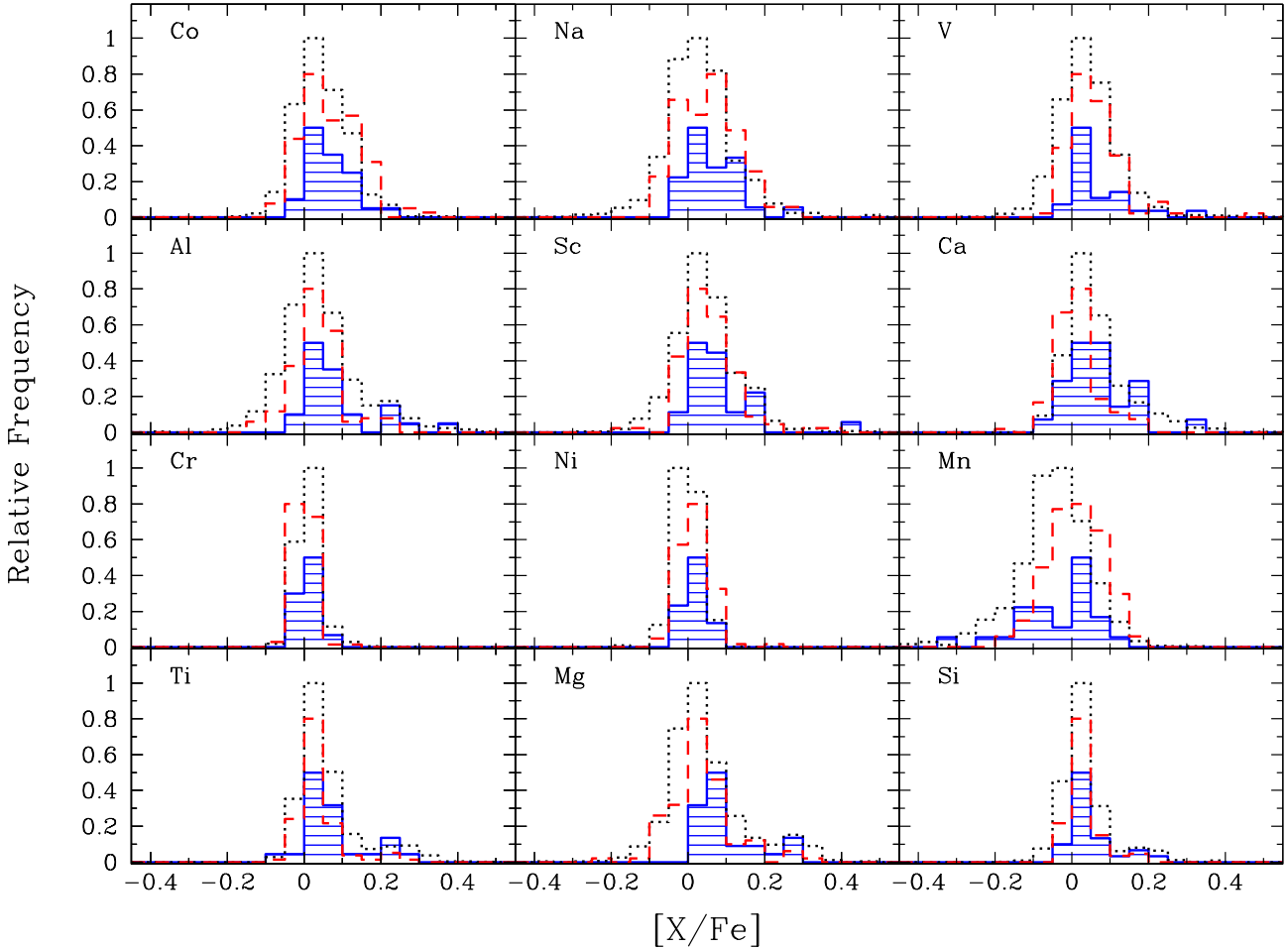}\caption{[X/Fe] distribution of the different elements. 
The stars with giant planets and without planets are represented by a red dashed and black dotted lines, respectively. 
The stars that exclusively host Neptunians and super-Earth planets are represented by a shaded blue.
The Neptunian/super-Earth and Jovian hosts distributions were set smaller (0.5 and 0.8 times, respectively) for the sake of clarity.}
\label{x_fe_hist}
\end{figure*}

\section{Abundances in planet hosts}

In this section we perform a detailed analysis of the [X/Fe] distributions of stars hosting Jupiter-mass planets, stars with Neptunes and super-Earths, and
stars without planets. At the end of this section we investigate the connection of metal-poor planet-hosting stars to the Galactic thick disc.

\begin{table*}
\centering
\caption{Average abundance ratios [X/Fe] for stars without planets, with giant planets, and stars that exclusively host Neptunians, along with their rms 
dispersion, the number of stars used in their determination, and the difference of averages between Neptunian and Jovian hosts and stars without planets.}
\label{xfe}
\begin{tabular}{llllllllllll}
\hline
\hline 
Species & \multicolumn{3}{c}{Jovian hosts} & \multicolumn{3}{c}{Neptunian hosts} & \multicolumn{3}{c}{Without planets} & \multicolumn{2}{c}{Difference of averages}\tabularnewline
X & $<$[X/Fe]$>$ & $\sigma$ & $N$ & $<$[X/Fe]$>$ & $\sigma$ & $N$ & $<$[X/Fe]$>$ & $\sigma$ & $N$ & Jovian - & Neptunian -\tabularnewline
 &  &  &  &  &  &  &  &  &  & Non-hosts & Non-hosts\tabularnewline
\hline 
Na  & 0.054 & 0.076 & 109 & 0.061 & 0.069 & 26 & 0.031 & 0.098 & 975 & 0.023 & 0.030\tabularnewline
Mg  & 0.031 & 0.083 & 109 & 0.100 & 0.083 & 26 & 0.048 & 0.103 & 976 & -0.017 & 0.052\tabularnewline
Al  & 0.037 & 0.066 & 109 & 0.084 & 0.097 & 26 & 0.044 & 0.104 & 969 & -0.007 & 0.040\tabularnewline
Si  & 0.030 & 0.041 & 109 & 0.050 & 0.059 & 26 & 0.034 & 0.066 & 976 & -0.004 & 0.016\tabularnewline
Ca  & 0.013 & 0.058 & 109 & 0.075 & 0.082 & 26 & 0.062 & 0.083 & 976 & -0.049 & 0.013\tabularnewline
Sc  & 0.056 & 0.073 & 109 & 0.085 & 0.060 & 26 & 0.047 & 0.085 & 947 & 0.009 & 0.038\tabularnewline
Ti  & 0.037 & 0.055 & 109 & 0.075 & 0.090 & 26 & 0.062 & 0.085 & 976 & -0.025 & 0.013\tabularnewline
V  & 0.059 & 0.075 & 109 & 0.064 & 0.076 & 26 & 0.041 & 0.084 & 973 & 0.018 & 0.023\tabularnewline
Cr  & -0.003 & 0.023 & 109 & 0.009 & 0.027 & 26 & 0.014 & 0.039 & 976 & -0.017 & -0.005\tabularnewline
Mn  & 0.011 & 0.073 & 109 & -0.043 & 0.100 & 26 & -0.047 & 0.096 & 976 & 0.058 & 0.004\tabularnewline
Co  & 0.065 & 0.073 & 109 & 0.066 & 0.054 & 26 & 0.048 & 0.078 & 975 & 0.017 & 0.018\tabularnewline
Ni  & 0.019 & 0.042 & 109 & 0.019 & 0.028 & 26 & 0.000 & 0.039 & 976 & 0.019 & 0.019\tabularnewline

\hline 
\end{tabular}
\end{table*}

\subsection{[X/Fe] of planet host stars}

Through studing the [X/Fe] histograms of exoplanet-hosting starsone can characterize the distribution of the individual elements relative to iron for the 
sample in general, and find the limits of the distributions.
The [X/Fe] distributions and the corresponding cumulative fractions for the total sample are depicted in Fig.~\ref{x_fe_hist} and Fig.~\ref{ks}, respectively. 
In the plots (and in the subsequent figures) we used the average of TiI \& TiII  for Ti, the average of CrI \& CrII for 
Cr, and the average of ScI \& ScII for Sc to increase the statistics. The abundance ratios do not change when using the mean values as compared to 
using the different ions separately.

The inspection of Fig.~\ref{x_fe_hist} and Fig.~\ref{ks}
suggests that, in general, the [X/Fe] distributions of Neptunian hosts (NH) start at higher [X/Fe] values compared to the distributions of their giant 
planet-hosting counterparts. For most elements this shift in part reflects the fact that Neptunian and Jovian hosts (JH) have a different metallicity distribution 
and hence different [X/Fe] ratios because of the Galactic chemical evolution trends. However, this is not the case of Mg, for which the shift of the distribution 
toward the higher [X/Fe] values is most evident. This shift exists even when we compare the two subsamples in the same metallicity interval.
The Kolmogorov-Smirnov (K--S) statistics predict a $10^{-4}$ probability (P) that Neptunian and Jovian host stars have the same [Mg/Fe] distribution.
The K--S tests also give significant probabilities that the [X/Fe] distribution of the two planet-hosting samples are different for Ca (K--S P $\approx$ $10^{-4}$) , 
Ti (K--S P $\approx$ 0.04),  and Mn (K--S P $\approx$ 0.05). 

Table~\ref{xfe} lists the average values of [X/Fe] for three groups of stars, along with their $\emph{rms}$
dispersion, the number of stars used in their determination, and the difference of averages between Neptunian and Jovian hosts
and stars without planets. These differences range from about -0.005 (Cr) to 0.052 (Mg) for NHs. 
The maximum overabundance of JHs are found for manganese $\Delta$[Mn/Fe] $\approx$ 0.06 dex.
We note that previous studies (Bodaghee et al. \cite{Bodaghee-03}; Zhao et al. \cite{Zhao-02}) 
have also suggested the difference in [Mn/Fe] between PHSs and comparison stars, though the difference was statistically within their scatter. 
As mentioned above, Kang et al. (\cite{Kang-11}) also claimed  to have found even significant overabundances of [Mn/Fe] of planet-hosting stars 
with the K--S probability 0.0015\% that the host and non-host stars have the same [Mn/Fe] distribution.
We note that these differences can be partially ``affected`` by the underlying [Fe/H] distributions of planets with different mass and control sample.

To evaluate the probability that the abundances of the two planet-hosting groups and the sample without planets have the same [X/Fe] distribution, we performed a K--S test.
To avoid confusion from the multiple values of [X/Fe] (for ''$\alpha$-like`` elements) at low metallicities due to the thin and thick disk 
(e.g. Bensby et al \cite{Bensby-03}; Adibekyan et al. \cite{Adibekyan-11}) and to remove (at least partially) 
the differences of the underlying iron distributions for the three samples, we established a cutoff in [Fe/H].
Only stars with metallicities between -0.2 and +0.4 dex were used for the K--S tests. The values of the probability that the [X/Fe] distributions belong to 
the same population (Prob) are presented in the Table~\ref{ks}. 
We note that, in general, the Prob values did not change when we extended the metallicity region from +0.4 to +0.55 dex (+0.55 dex is the metallicity 
limit for our sample), but they decreased (especially for ''$\alpha$-like`` elements) if we added the stars with metallicities from -0.2 to -0.6 dex 
(-0.6 dex is the lower metallicity limit for planet hosts in our sample). 
This is expected because in the low-metallicity regime almost all planet hosts have high [$\alpha$/Fe] values, typical for thick-disk stars, 
while stars without planets belong to the thin and thick disks with low and high [$\alpha$/Fe] values, respectively (see Sect 3.2 and 3.3).
For the metallicity distribution, the probability that the [Fe/H]
distributions of NHs and non-hosts belong to the same population is about 99\% and for JHs and non-hosts the test gives 0\% probability. The K--S probabilities for NHs 
reflects the real differences in [X/Fe] between these and non-host stars because their underlying iron distributions fully match.
The K--S test implies that the stars hosting Neptunians and stars without planets are indeed from separate groups in terms of [Mg/Fe], [Si/Fe], [Sc/Fe] and [Al/Fe].
For Jupiter-mass planet hosts the K--S test implies that they are different from stars without planets by their [X/Fe] distribution for all the elements (except Ti and V)
although we should note that their [Fe/H] distributions are different as well.

\begin{figure}
\centering
\includegraphics[angle=270,width=1\linewidth]{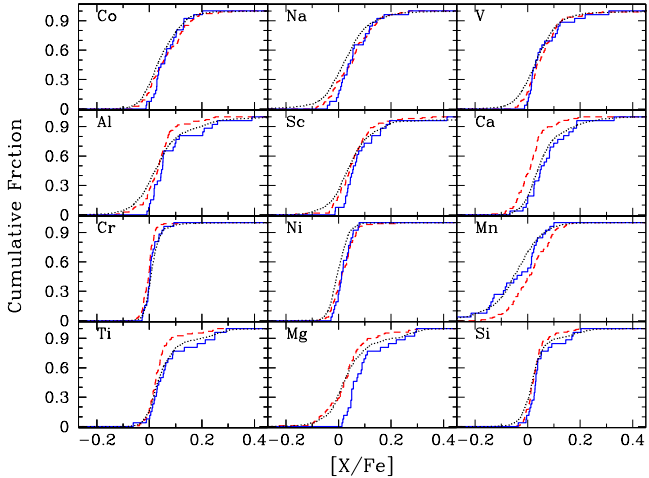}\caption{Cumulative distribution functions of [X/Fe]. 
The stars with giant- and low-mass planets are represented by a red dashed and a blue solid line, respectively. 
The stars without planets are represented by a black dotted line.}
\label{ks}
\end{figure}

\begin{figure}
\centering
\includegraphics[width=1\linewidth]{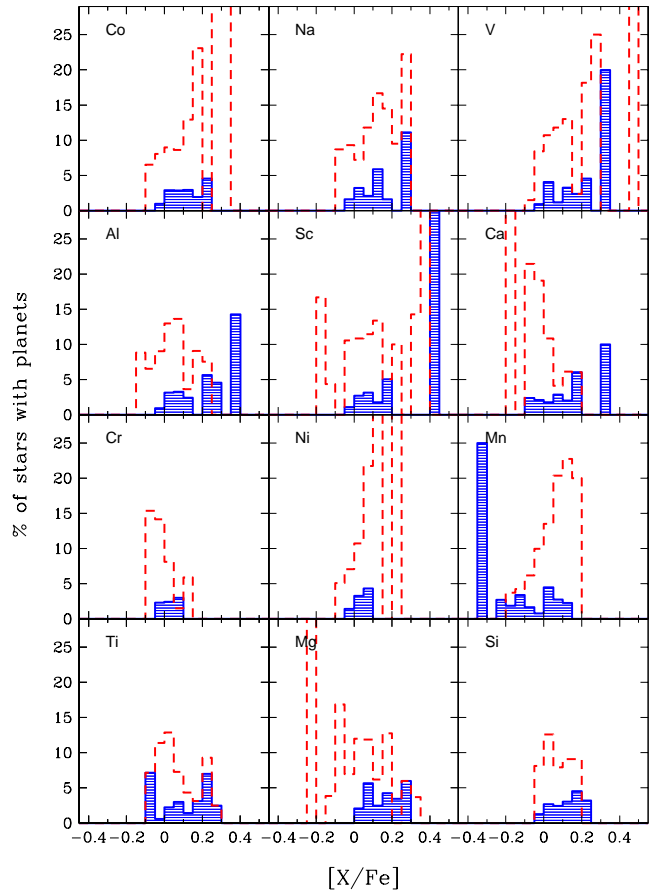}\caption{Percentage of stars with giant (red dashed) and exclusively Neptunian and super-Earth 
(shaded blue) planets as a function of [X/Fe]. For the sake of clarity we cut the y axis at 30 \%.}
\label{x_fe_freq}
\end{figure}


\begin{figure*}
\centering
\includegraphics[width=0.8\linewidth]{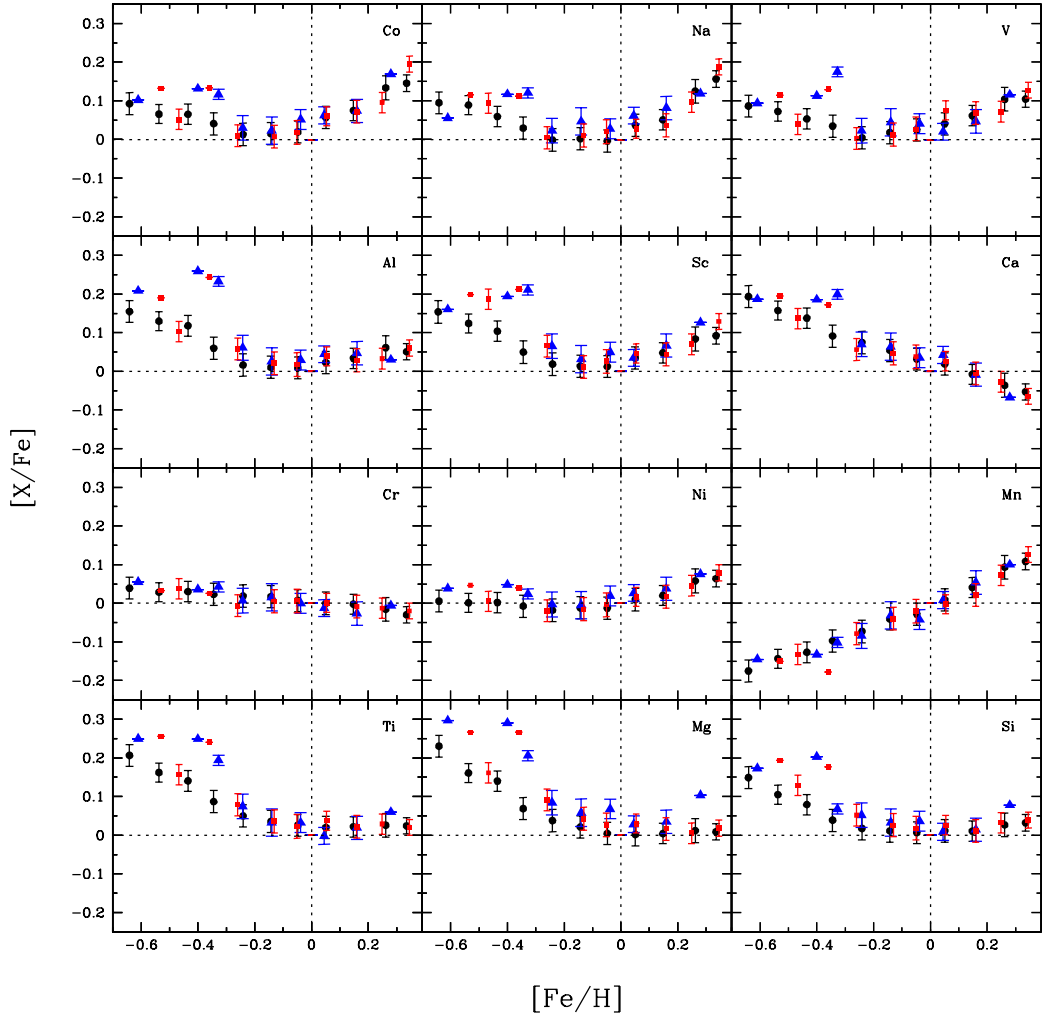}\caption{[X/Fe] abundance ratios against [Fe/H] for the stars with and without planets. The symbols and error bars 
indicate the average and standard deviation, respectively, of each bin (0.1 dex). The red squares and blue triangles represent stars 
with Jupiter-mass and Neptunian/super-Earth mass planets, respectively. The black circles refer to the stars without a
planetary companion. The black dashed lines target the solar value.}
\label{x_fe}
\end{figure*}

Fig.~\ref{x_fe_freq} illustrates the [X/Fe] distribution from a different perspective: the histograms of the number of stars with  giant (red dashed) and low-mass planets
(shaded blue) compared to the total number of stars of each bin (0.05 dex). We note that some bins with extreme [X/Fe] values show an unreal 100\% percentage 
with only one star in the bin. For the sake of clarity we cut the y axis at 30\%. For Co, Na, Ni, V, and Mn we observe that there is a general 
increase in the percentage of stars with giant planets, with increasing [X/Fe]. An opposite trend shows the distributions of Ca and Cr. It is difficult to 
claim that there is an opposite trend for Mg because the distribution is in general flat when we exclude the lowest [Mg/Fe] bin with the ''unreal`` 100\% frequency.
We also found that the probability to find a Neptunian mass planet is increasing if the host has high [X/Fe] values for all studied elements except Mn.
Interestingly, most of the NH stars have [Al/Fe], [Sc/Fe] and particularly [Mg/Fe] values higher than 0, although JHs are also observed  at
much lower [X/Fe] values. We again note that the number of stars that exclusively host  Neptunes and super-Earth planets is small, therefore, the conclusions 
regarding them should be considered with caution.

\begin{figure*}
\begin{center}$
\begin{tabular}{cc}
\includegraphics[width=0.45\linewidth]{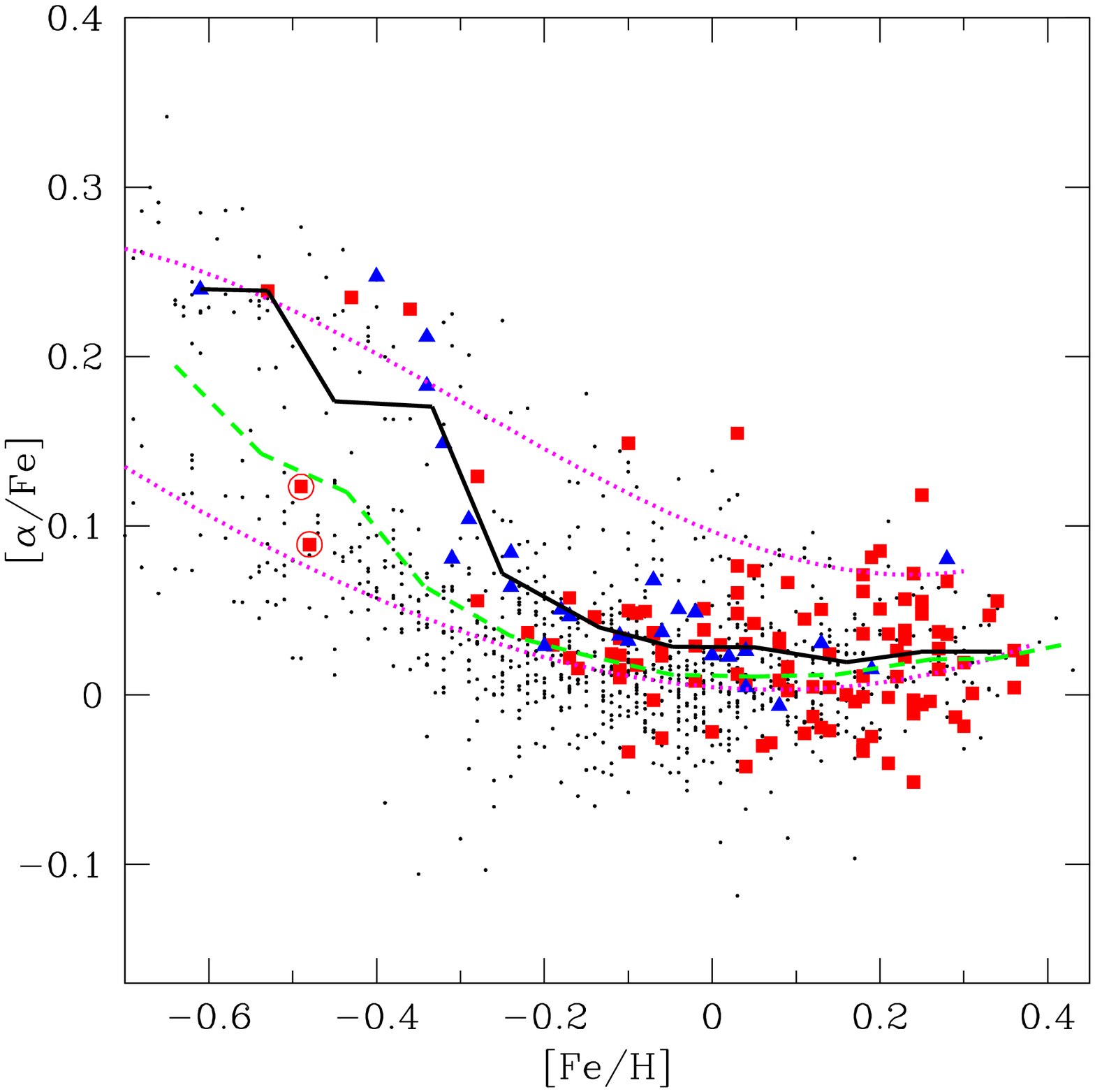}&
\includegraphics[width=0.45\linewidth]{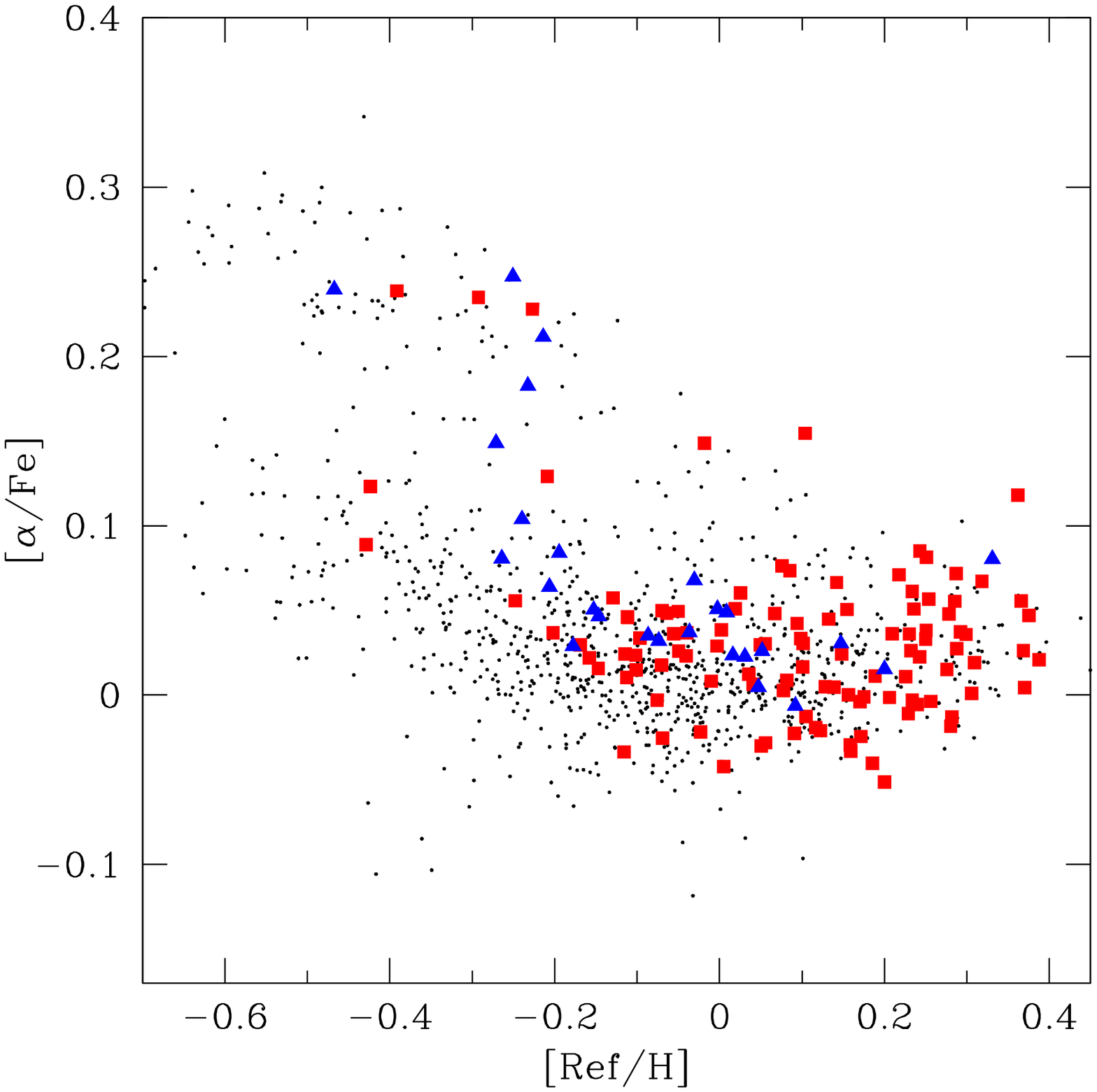}
\end{tabular}$
\end{center}
\caption{[$\alpha$/Fe] versus [Fe/H] ($\emph{left}$) and [$\alpha$/Fe] versus [Ref/Fe] ($\emph{right}$) for the total sample. 
The red squares refer to the Jovian hosts and the blue triangles  refer to the stars hosting exclusively Neptunians and super-Earths. The black dots represent
the stars without planets. The black solid and green dashed lines represent the mean distributions of the planet host and non-host samples, respectively.
The magenta dotted lines are the cubic fits for the low-$\alpha$ (thin disc) and the high-$\alpha$ (thick disk + high-$\alpha$ metal-rich)
stars. The two metal-poor Jovian hosts in the red circles are the stars with low $\log\,g$ values discussed in $\emph{Sect. 3.2}$.}
\label{alpha_fe}
\end{figure*}
\subsection{[X/Fe] versus [Fe/H]}

Previous planet-host studies (e.g. Neves et al. \cite{Neves-09}; Petigura \& Marcy \cite{Petigura-11}; Kang et al. \cite{Kang-11}) 
have found that, in addition to iron, the abundances of various other metals are enhanced in these stars compared to stars with no known planets.
Given the metal-rich nature of planet-hosting stars (higher overall iron content), they are indeed expected to have a higher content of 
other metals as well. Some authors also referred to the existence of different [X/Fe] trends in planet-hosting stars compared to 
stars without planets for the same metallicity. For instance, potential differences were detected by Sadakane et al. (\cite{Sadakane-02}) for vanadium
and cobalt, by Bodaghee et al. (\cite{Bodaghee-03}) for Ti, Mn, V, and Co, and by Gilli et al. (\cite{Gilli-06}) for V, Co, Mg, and Al. Robinson et al. 
(\cite{Robinson-06}) claimed to have found significant overabundances of [Si/Fe] and [Ni/Fe]. Gonzalez \& Laws (\cite{Gonzalez-07}) also found that 
Al and Si abundances are systematically lower for the planet-hosting stars in the higher metallicity region, and that the Ti abundance exhibits 
the opposite trend, which implies that the abundances of Na, Mg, Sc, and Ni might have some differences between planet hosts and stars without planets.
Recently, Kang et al. (\cite{Kang-11}), studying 34 PHSs and 18 comparison G-type stars, found that the [Mn/Fe] ratios of planet-hosting stars are higher than 
those of comparison stars over the entire metallicity range. These authors also found that [X/Fe] ratios of Mg, Al, Sc, Ti, V, and Co for PHSs are higher than those 
of comparison stars in metal-poor stars of [Fe/H] $<$ -0.4 by more than 0.2 dex (they had only two stars with and two stars without planets in the mentioned region).
Most of these studies were in general limited to small samples of planet-hosting and stars without planets and as a result the general picture is so very difficult 
and contradictory.

Knowing that different elements show different trends of abundances in the metal-rich and metal-poor domain
in the Galactic chemical evolution, studing the [X/Fe] histograms is not sufficient to find small/subtle abundance differences between stars 
harboring planets with different masses and stars without planets. To explore the importance of individual elements in planet formation it
is necessary to investigate the elemental abundances along with the metallicity. In this subsection, we will analyze whether there are any differences 
in the abundances of stars with and without planets for the same value of [Fe/H].

Fig.~\ref{x_fe} shows the averages and standard deviations of [X/Fe] ratios for each [Fe/H] bin. The sizes of the bins are 0.1 dex. The red squares
and blue triangles represents stars with Jovian-mass and Neptunian-mass planets, respectively. The black circles refer to the stars without
planetary companion. As shown in the figure, planet-hosting stars, both hosting Neptunes and Jupiters, are more abundant in $\alpha$-elements than
comparison stars when the metallicity is low. Giant-planet hosts have higher [Sc/Fe] and [Ti/Fe] values  than stars without planets when [Fe/H] $<$ -0.2 dex, 
higher [Al/Fe] and [Si/Fe]  when [Fe/H] $<$ -0.1 dex, and higher [Mg/Fe] ratios when [Fe/H] $<$ 0.1 dex.   We note that the [X/Fe] values in 
the -0.5 $<$ [Fe/H] $<$ -0.4 dex
metallicity bin for the Jupiter-hosting stars are not fitted with the general [X/Fe] vs [Fe/H] trends. In that bin we have three stars, two of which have ``thin'' disk [X/Fe] 
abundance ratios. Interestingly, these two stars are the most ``evolved'' planet hosts of our sample in that iron-poor regime. 
The stars, HD171028 and HD190984, have 3.84 and 4.02 dex $\log\,g$ values, respectively, although the average $\log\,g$ of the metal-poor ([Fe/H] $<$ -0.2 dex)
planet hosts is 4.35 dex, and for the total planet-hosting sample it is 4.37 dex. 
It is difficult to conclude why these two stars show 
``peculiar'' $\alpha$-element abundances. Increasing the statistics will help to resolve this uncertainty.

\begin{table}
\centering
\caption{The K--S probabilities that stars with planets of different mass and non-planet host stars have the same [X/Fe] distribution. Only stars with 
-0.2 $<$ [Fe/H] $<$ 0.4 dex are considered.}
\label{ks}
\begin{tabular}{lcc}
\hline
\hline 
Species & {Jovian hosts} & {Neptunian hosts}\tabularnewline
& K--S Prob &K--S Prob \tabularnewline
\hline 
Na  & 0.004 & 0.242\tabularnewline
Mg  & 0.010 & 0.000\tabularnewline
Al  & 0.009 & 0.045\tabularnewline
Si  & 0.012 & 0.014\tabularnewline
Ca  & 0.002 & 0.328\tabularnewline
Sc  & 0.049 & 0.018\tabularnewline
Ti  & 0.629 & 0.763\tabularnewline
V  & 0.123 & 0.519\tabularnewline
Cr  & 0.049 & 0.099\tabularnewline
Mn  & 0.002 & 0.372\tabularnewline
Co  & 0.031 & 0.142\tabularnewline
Ni  & 0.000 & 0.195\tabularnewline
Fe  & 0.000 & 0.990\tabularnewline
\hline 
\end{tabular}
\end{table}

For low-mass planet host the increase in [X/Fe] ratios starts at higher metallicities than for high-mass planet hosts. Stars hosting Neptune-like planets have higher 
[Ti/Fe] values than their non-host counterparts when [Fe/H] $<$ -0.2 dex, higher [Al/Fe], [Si/Fe], [Co/Fe], [V/Fe] and [Sc/Fe] when [Fe/H] $<$ 0.0 dex, and they have higher [Mg/Fe] 
and [Na/Fe] (for Na the difference is smaller) abundance ratios in all bins (except the most metal-poor and metal-rich bins where we have only one star with 
low [Na/Fe] ratio).
Low-mass planet hosts also exhibit slightly higher [Ni/Fe] abundance ratios, although they are still consistent within the error bars in all metallicity bins except 
the -0.2 $<$ [Fe/H] $<$ -0.1 dex.
With our relatively large and homogeneous sample we did not observe any abundance differences in manganese between stars with and without planets, as claimed in 
Kang et al. (\cite{Kang-11}), and we observed no nickel enrichment in stars with giant planets, as reported in Robinson et al. (\cite{Robinson-06}).

Based on these results, we can assume that some metals other than iron are involved in the process of planet formation, especially when the amount of iron is lower than solar.
Iron is not the only abundant refractory element in the solar system. There are other fairly abundant elements (e.g. Mg, Si) with condensation temperatures comparable to iron 
(Lodders \cite{Lodders-03}; Lodders et al. \cite{Lodders-09}) that are very important contributors to the composition of dust in planet-forming regions 
and represent the principal components of rocky-type planets. Thus, these results are consistent with the expectations, since a high [X/Fe] ratio means higher 
``global metallicity'' and are moreover supported by the theoretical studies using the core-accretion model (e.g. Ida \& Lin \cite{Ida-05}; Mordasini et al. \cite{Mordasini-12}).
However, our analysis indicates that planet-hosting stars have higher [X/Fe] values than stars without planets for elements that are less abundant (e.g. Sc, Ti). This may indicate
that not all solids are equally effective for planet formation. Alternatively, some elements (like Si or Mg) may ``really`` stimulate planet formation processes, although
observed overabundance of otherelements may just reflect the similarities of the Galactic chemical evolution of the elements. Particularly, Si, Mg, Al, Sc, and Ti show similar 
[X/Fe] trends with [Fe/H], although in this case one might expect overabundance of Ca in planet host stars, which is not observed. 
More analyse are needed to understand which elements show a secondary correlation with exoplanet presence.

Although the observed abundance differences between the stars with and without planets are in general tiny compared to the standard deviations of the samples, they are 
systematic, and we believe that increasing the sample of planet-hosting stars will yield statistically more significant results.

\subsection{Exoplanets and the thick disc}

Haywood (\cite{Haywood-08,Haywood-09}) reported that at metallicities [Fe/H] $<$ -0.3 most stars known to harbor giant planets belong to the thick
disk rather than to the thin disc. Gonzalez (\cite{Gonzalez-09}), using three samples of nearby stars hosting giant 
planets, showed that if compare stars with planets according to mass abundance of the refractory elements important for planet formation 
(instead of only iron), then thick disk and [Fe/H]-poor thin disk stars with planets have similar distributions. Studing this 
connection of planet-hosting stars with the thick disk is the subject of this subsection.

In Fig.~\ref{alpha_fe} we plot [$\alpha$/Fe] against [Fe/H] and [$\alpha$/Fe] against [Ref/Fe] for the total sample. 
The ``Ref'' index is the one proposed by Gonzalez (\cite{Gonzalez-09}) and quantifies the mass abundances of Mg, Si and Fe.
The red squares refer to the Jovian hosts and the blue triangles refer to the stars exclusively host Neptunians and
super-Earths. The black dots represent the stars without planets.
To illustrate the path of planet formation along the abundance ratios [$\alpha$/Fe] we present the mean of the distribution (black solid line).
For the comparison the mean of the [$\alpha$/Fe] distribution of the stats without planets are also presented  (green dashed line).
The magenta dotted lines are the cubic polynomial fits for the low-$\alpha$ (thin disc) and the high-$\alpha$ stars
(thick disk + high-$\alpha$ metal-rich, see Adibekyan et al. (\cite {Adibekyan-11}) for the chemical separation of the sample).

With this data we confirm the finding by Haywood (\cite{Haywood-08,Haywood-09}) that giant planet incidence is greater among the thick-disk population
than among the thin-disk population for [Fe/H] $<$ -0.3 dex. We also show that this can be extended to the Neptunian hosts. 
At the mentioned metallicity region we have only three stars that
harbor a planet in the thin disk (two of them are the aforementioned planet hosts with low $\log\,g$ values) and eight planet hosts in the thick disc%
\footnote{The separations between the Galactic thin and thick discs were taken from Adibekyan et al. (\cite{Adibekyan-11}).%
}. 
Interestingly, this thin/thick proportion changes dramatically when we apply a purely kinematic approach to separate the thin and thick discs.
The kinematical separation suggests that  six or three host are from the thick disc, four or six stars belong to thin disk and one or two stars can be classified as 
transition stars depending on the kinematic criteria 
used - Bensby et al (\cite{Bensby-03}) or Reddy et al. (\cite{Reddy-06}), respectively (see also Adibekyan et al. \cite{Adibekyan-11}).
Moreover, following Gonzalez \cite{Gonzalez-09}), and considering the “Ref” index  instead of [Fe/H], we can see that the observed high abundance of metal-poor PHSs in the 
thick disk ``disappears'': most metal-poor planet-hosting stars in the thick disk have the same [Ref/H] distribution as their thin-disk counterparts 
(see the \emph{right} panel of Fig.~\ref{alpha_fe}). This indicates that the thick-disk planet hosts might have less Fe than thin-disk hosts
but they have relatively more Mg and Si, so they are not as metal -poor in terms of their ability to form planets.

In the Fe-poor regime it is difficult to conclude abouts the main reason for most of the planet-hosting stars to lie in the high-$\alpha$/thick-disk region. 
Generally, the Galactic thick disk is composed of relatively old stars (e.g. Bensby et al. \cite{Bensby-05}; Adibekyan et al. \cite{Adibekyan-11}) that move in 
Galactic orbits with a large-scale height and long-scale length (e.g. Robin et al. \cite{Robin-96}; Juri\'{c} et al. \cite{Juric-08}). At the same time, the thick-disc
stars are known to have higher [$\alpha$/Fe] ratios than the stars from thin disk at the same [Fe/H]. Below we present two indications that impel us to suppose
that the certain chemical composition is the decisive factor, not the Galactic birth radius.

\begin{figure}
\centering
\includegraphics[angle=270,width=1\linewidth]{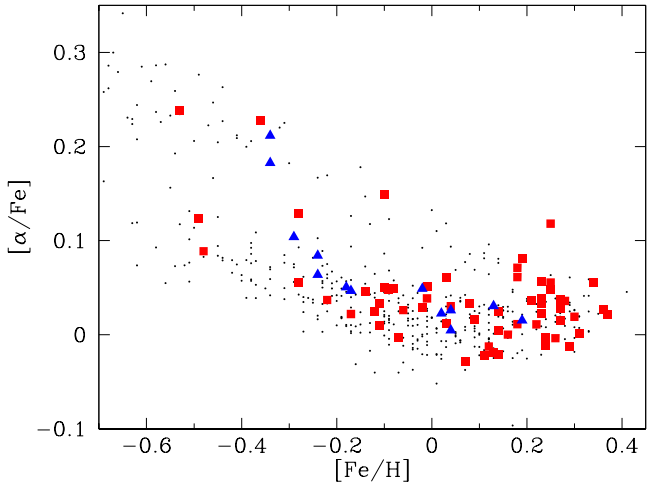}\caption{The same as Fig.~\ref{alpha_fe} (\emph{left} panel), but for stars with 
\emph{$T{}_{\mathrm{eff}}$} =\emph{$T{}_{\mathrm{\odot}}$} $\pm$300 K.}
\label{alpha_fe_300k}
\end{figure}

\begin{figure}
\centering
\includegraphics[width=1\linewidth]{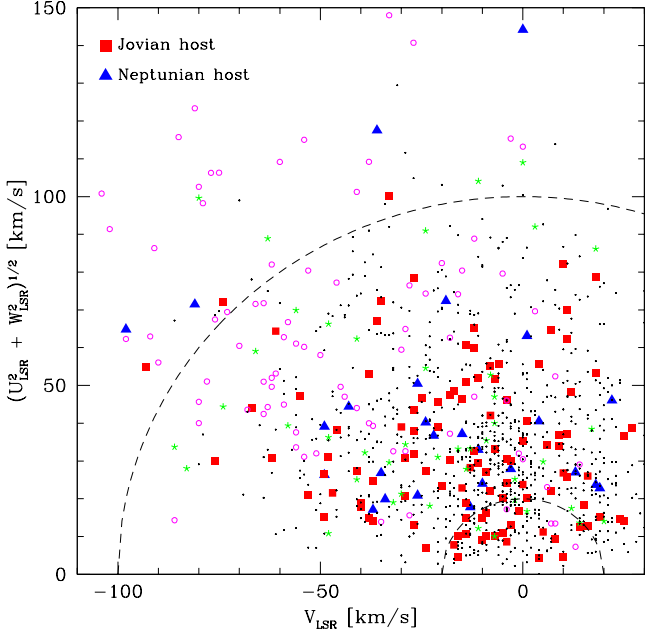}\caption{Toomre diagram for the planethosting and stars without planets. The red squares and blue 
triangles represent stars with Jupiter- and Neptune-mass planets, respectively. The magenta circles and green asterisks refer to the high-$\alpha$ 
metal-poor (chemically defined thick disc) and high-$\alpha$ metal-rich stars without planets, and the black dots refer to the chemically defined thin disk non-host stars.
Dotted lines indicate constant peculiar space velocities, $\upsilon_{pec} = (U^2_{LSR} + V^2_{LSR} + W^2_{LSR})^{1/2} = 20$ and 100 km $s^{-1}$.}
\label{fig-13}
\end{figure}

$\bullet$ Fig.~\ref{alpha_fe} clearly shows that planet-hosting stars show a continuous increase in [$\alpha$/Fe] with decreasing [Fe/H] at metallicities from -0.2 to -0.3 dex 
(starting from the thin disc, they rise to the thick disc) while the thin and thick disk stars without planets are separated very well by their 
[$\alpha$/Fe] ratios (see Adibekyan et al. \cite{Adibekyan-11}). Fig.~\ref{alpha_fe_300k} better illustrates this separation; we plot only stars with
effective temperatures close to the Sun by $\pm$300 K. In the metalicity region from -0.4 to -0.2 dex there is a clear separation between the two Galactic discs and 
there is no ``transition`` population between them, while the rise of [$\alpha$/Fe] for planet hosts still exists. 
This fact may be an indication that planet-hosting stars follow a certain chemical evolution trend that requires a certain chemical composition for their formation.
This does not mean that the planet-hosting stars constitute a separate stellar population, and only suggests that perhaps a planet can be formed everywhere the 
chemical requirements are satisfied.

$\bullet$ The second hint, which may also indicate the importance of the $\alpha$ elements, is that planet-hosting stars start to have high [$\alpha$/Fe] ratios 
at lower metallicites when they still belong to the thin disc. This is illustrated in the $\emph{left}$ panel of Fig.~\ref{alpha_fe} where one can see that
the average [$\alpha$/Fe] ratio for planet-hosting stars (black line) is higher than that for stars without planets (green dotted line) when [Fe/H] $\lesssim$ 0 - 0.1 dex.

In this subsection we considered low-mass and high-mass planet hosts together because, in general, they show similar [$\alpha$/Fe] trends with the metallicity 
(although at ''higher`` metallicities most of the Neptunian hosts lie above the ''average`` distribution of the combined host sample).

The fact that most of the metal-poor planet hosts belong to the thick disk leads to an interesting conclusion. This suggests that when we calculate the 
frequency of planet hosts at low metallicities we should consider only stars belonging to the same Galactic population.
In particular, the frequency of planet hosts in the metallicity region from -0.6 to -0.3 dex
is about 5\%, but when we consider only stars belonging to the thick disc, we will have about 14\% and in the thin disk about 2\%. To compare, 
 the frequency of planet hosts in the metallicity region from -0.3 to 0 dex in the thin disk is about 9\%. 
This may also suggest that the observed paucity of very metal-poor planet hosts 
just reflects the low density of the thick disk stars in the solar neighborhood.

\begin{table}
\centering
\caption{The average values of the Galactic orbital parameters and ages for the stars with giant planets, stars that exclusively host Neptunians,
stars without planets, and stars without planets with low-$\upsilon_{pec}$. The standard errors of the mean values are also presented.}
\label{table-gal}
\begin{tabular}{lcccc}
\hline
\hline
 & {\tiny Z$_{\mathrm{max}}$} & {\tiny Ecc} & {\tiny R$_{Gal}$} & {\tiny Age}\tabularnewline
\hline 
{\tiny NHs} & {\tiny 0.33$\pm$0.06} & {\tiny 0.19$\pm$0.02} & {\tiny 8.00$\pm$0.001} & {\tiny 5.5$\pm$0.3}\tabularnewline
{\tiny JHs} & {\tiny 0.21$\pm$0.02} & {\tiny 0.14$\pm$0.01} & {\tiny 7.99$\pm$0.002} & {\tiny 4.8$\pm$0.3}\tabularnewline
{\tiny Non-hosts} & {\tiny 0.29$\pm$0.02} & {\tiny 0.15$\pm$0.00} & {\tiny 7.99$\pm$0.001} & {\tiny 5.2$\pm$0.1}\tabularnewline
{\tiny low-$\upsilon_{pec}$} & {\tiny 0.10$\pm$0.01} & {\tiny 0.05$\pm$0.00} & {\tiny 7.98$\pm$0.003} & {\tiny 4.8$\pm$0.3}\tabularnewline
\hline 
\end{tabular}
\end{table}

%
\begin{figure*}
\centering
\includegraphics[angle=270,width=0.75\linewidth]{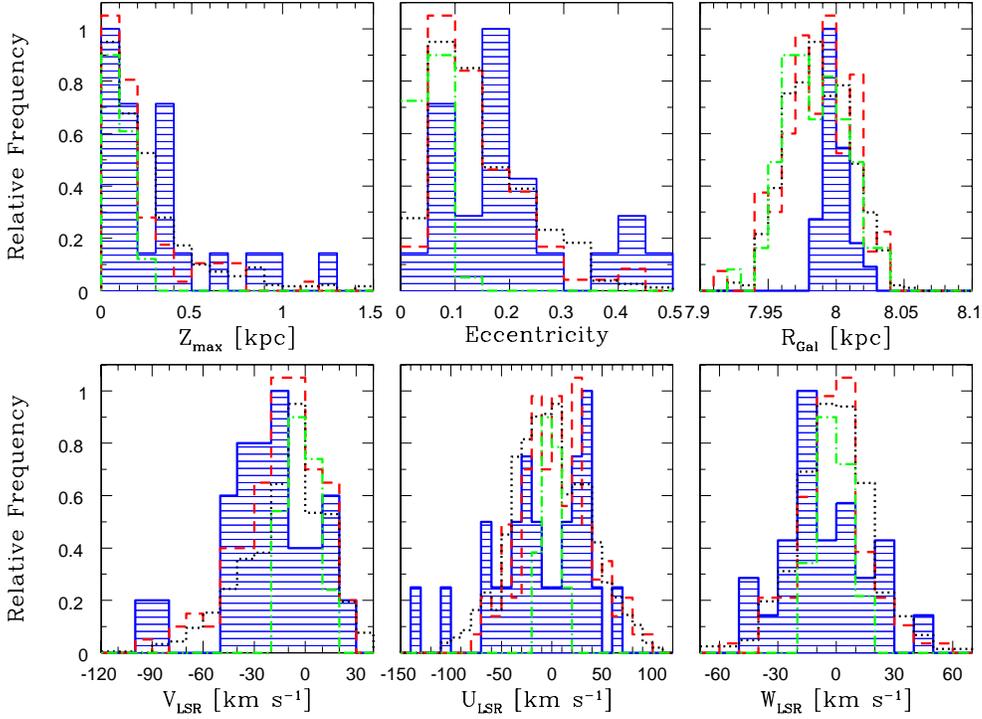}\caption{Distributions of the Galactic orbital parameters and space velocities of 
the stars with Neptune-mass planets (shaded blue), with Jupiter-mass planets (red dashed line), stars without planets and [Fe/H] $>$ -0.6 dex (black dotted line), 
and stars without planets with [Fe/H] $>$ -0.6 dex and $\upsilon_{pec} < 20$ km $s^{-1}$ (green dotted-dashed line). The distributions of different groups were set lower/higher
for the sake of clarity.}
\label{fig-14}
\end{figure*}

\section{Kinematics of planet-hosting stars}

So far,  more than a dozen studies were published focused on the kinematics of the PHSs and their relation to different stellar populations. 
The first papers did not find any significant kinematic peculiarity of PHSs (e.g. Gonzalez \cite{Gonzalez-99}; Reid \cite{Reid-02};
Barbieri \& Gratton \cite{Barbieri-02}). 
Conversely, Haywood (\cite{Haywood-08};\cite{Haywood-09}), combining the chemical and kinematic properties of the PHSs, concluded that most metal-rich 
stars that host giant  planets originate from the inner Galactic disk. 
The same scenario for the origin of metal-rich planet hosts were proposed in some other papers 
(e.g. Ecuvillon et al. \cite{Ecuvillon-07}; Santos et al. \cite{Santos-08}; Viana Almeida et al. \cite{Almeida-09}).

To study the kinematics of planet-hosting stars in Fig.~\ref{fig-13} we presented their positions in the \emph{UW} vs \emph{V} velocities space (Toomre diagram).
For details of the computation of the space velocity components we refer the reader to Adibekyan et al. (2012, submitted).
In the diagram we used only stars without planets with [Fe/H] $>$ -0.6 dex, because  we found no planet below this value.
Dotted lines indicate constant peculiar space velocities, $\upsilon_{pec} = (U^2_{LSR} + V^2_{LSR} + W^2_{LSR})^{1/2} = 20$ and 100 km $s^{-1}$.
We first note is that  there is no kinematically very ``cool'' Neptunian host with $\upsilon_{pec} < 20$ km $s^{-1}$, while 18 stars that
host Jovian planets ($\sim$17 \%) have peculiar space velocities slower than 20 km $s^{-1}$. The number of stars without planets and 
$\upsilon_{pec} < 20$ km $s^{-1}$ is 121, which is $\approx$13 \% of the total sample of non-hosts ([Fe/H] $>$ -0.6 dex).

First we checked if the different groups lie at the same distances from the Sun. We found that NHs are, on average, closer to the Sun 
than JHs or stars without any planet companions.
The average distance of NHs is $\approx$18 pc, while the average distance of JHs is $\approx$32 pc, for stars without planets it is $\approx$38 pc, 
and for the stars with $\upsilon_{pec} < 20$ km $s^{-1}$ is $\approx$39 pc. This result is expected, because low-mass planets are easier to find at 
smaller distances due to the higher apparent magnitudes of their hosts. However, we have found that the difference in the distances is not the reason for the 
''high`` peculiar velocities of NHs. Among the stars without planets that are closest to the Sun (the average distance of this subsample was $\approx$18 pc) 
we found 18 stars ($\approx$16\% of the the total subsample) with $\upsilon_{pec} < 20$ km $s^{-1}$.

To understand the root of this kinematical ``peculiarity`` of NH stars we collected the main orbital and kinematical properties of 
these stars and compared them to those from other groups of stars. To obtain the Galactic orbital parameters of the stars, we cross-matched our sample with 
the Geneva-Copenhagen Survey (GCS) sample (Casagrande et al. \cite{Casagrande}), which provides the eccentricities of the orbits, the maximum vertical 
distance (Z$_{\mathrm{max}}$) a star can reach above the Galactic plane, the Galactic radial positions, and the ages of about 700 
of our stars 
(23 NHs, 77 JHs, 548 non-host stars with [Fe/H] $>$ -0.6 dex, and 55 non-hosts with [Fe/H] $<$ -0.6 dex).
The distributions of the Galactic orbital parameters and space velocities of the stars with low-mass planets (blue shade), with high-mass planets
(red line), stars without planets with [Fe/H] $>$ -0.6 dex (black line), and the non-host stars with [Fe/H] $>$ -0.6 dex and 
$\upsilon_{pec} < 20$ km $s^{-1}$ (green dotted line, hereafter low-$\upsilon_{pec}$ stars) 
are presented in Fig.~\ref{fig-14}. The average values of the physical parameters aand the ages for the groups of stars are presented in 
Table~\ref{table-gal}.

Inspection of the figure and the associated table shows that NH stars, on average, have higher Z$_{\mathrm{max}}$, higher eccentricities, and higher R$_{Gal}$ than their
high-mass planet-hosting counterparts and stars without planets. The differences, as expected, are higher when compared to the 
low-$\upsilon_{pec}$ stars.  From the bottom panels of Fig.~\ref{fig-14} we can see that NHs have lower $V_{\mathrm{LSR}}$ and W$_{\mathrm{LSR}}$ space velocity 
components than stars from other groups. All these differences indicate that NHs tend to belong to the ''thicker`` disc, 
although JHs show the same frequency in the ''thinner`` disk as the stars without planets in our sample. This finding confirms also the last column
of Table~\ref{table-gal} where one can see that NH stars are on average older than the JHs (although the difference is small).

It is difficult to conclude about the main reason of the observed tendency of NH stars to belong to the ''thicker`` disc. In Fig.~\ref{fig-13} we can see that NHs  
follow the distribution of high-$\alpha$ stars in the \emph{UW} vs \emph{V} velocity space. To see if the NHs and stars without planets with the same 
metallicity (-0.6 $<$ [Fe/H] $<$ 0.25) and kinematics ($\upsilon_{pec} > 20$ km $s^{-1}$) have the same abundances of $\alpha$-elements, we compared their
[Mg/Fe], [Si/Fe] and [Ti/Fe] ratios. We found that the NH stars are more enhanced by Mg (the difference in [Mg/Fe] $\approx$ 0.06 dex) than
by Si ($\Delta$[Si/Fe] $\approx$ 0.02 dex) and Ti ($\Delta$[Ti/Fe] $\approx$ 0.02 dex), and both groups of stars are enhanced by these elements compared
to the low-$\upsilon_{pec}$ stars without planets. This may be a subtle hint confirming our previous finding (see Sect. 3) that to form Neptune-mass 
planets, some elements, like Mg, are particularly important.

\section{Conclusions}

We presented a differential abundance analysis between planet-hosting and stars without planets for 12 refractory elements 
(Na, Mg, Al, Si, Ca, Ti, Cr, Ni, Co, Sc, Mn, and V) for a total of 1111 nearby FGK dwarf stars. Of these stars, 109  are known to
harbor high-mass planetary companions and 26 stars are  exclusively hosting Neptunians and super-Earths. The precise
spectroscopic parameters for the entire sample were taken from Sousa et al. (\cite{Sousa-08,Sousa-11a,Sousa-11b}) and the abundances were taken from
Adibekyan et al. (2012, submitted). 

The inspection of the [X/Fe] histograms suggests that in general, the [X/Fe] distributions of NHs start at higher [X/Fe] values
compared to the distributions of giant-planet hosts. This shift toward the higher [X/Fe]values is most clearly observed for Mg.
For Co, Na, Ni, V, and Mn we observe that there is a general increase in the frequency of stars with giant planets, with increasing [X/Fe].
We also found that most of the Neptunian hosts have [Al/Fe], [Sc/Fe] and [Mg/Fe] values higher than 0, although Jovian hosts can have 
much lower [X/Fe] values for these elements.

With our relatively large and homogeneous sample of planet-hosting and stars without planets we found that the [X/Fe] ratios of Mg, Al, Si, Sc, and Ti for giant hosts
are systematically higher than those of comparison stars  at [Fe/H] $\lesssim -0.1 \pm 0.1$ dex. Simultaneously, stars hosting Neptune-like planets
show higher [Ti/Fe] ([Fe/H] $<$ -0.2 dex),  [Si/Fe], [Al/Fe], [Co/Fe], [V/Fe] and [Sc/Fe] ([Fe/H] $<$ 0.0 dex), and higher [Mg/Fe] and [Na/Fe] (over the entire metallicity range)
values than their non-host counterparts. Low-mass planet hosts also exhibit slightly higher [Ni/Fe] abundance ratios, although they are still consistent within the error bars
 in all metallicity bins except the -0.2 $<$ [Fe/H] $<$ -0.1 dex.

We confirmed that planet incidence is greater among the thick-disk population than among the thin disk for [Fe/H] $<$ -0.3 dex. 
At lowest metallicities we observed only two stars with low [$\alpha$/Fe] ratios and both have lower
$\log\,g$ values (3.84 and 4.02 dex).

We also discussed recent debates about the high abundance of metal-poor planet hosts in the thick disk (Haywood \cite{Haywood-08,Haywood-09}; Gonzalez \cite{Gonzalez-09}), 
i.e. the main reason that most Fe-poor stars lie in the high-$\alpha$/thick-disk region - is it a special birth place in the Galaxy or a certain chemical composition.
Our results allow us to propose that the certain chemical composition and not the Galactic birth place is the determining factor.

The study of kinematical properties of planet host stars shows that there is no kinematically very ''cool`` low-mass planet host star with a peculiar
velocity slower than 20 km $s^{-1}$, while $\sim$17 \% of stars that host high-mass planets have $\upsilon_{pec} < 20$ km $s^{-1}$. Although we found that
NHs are closer to the Sun than JHs and stars without planets, this is not a reason of the lower velocity limit. Inspecting of the 
Galactic orbital parameters of these stars shows that NHs tend to belong to the ''thicker`` disk compared to their high-mass planet-hosting counterparts. 
We also found that they follow the distribution of high-$\alpha$ stars in the \emph{UW} vs \emph{V} velocity space, but they are more 
enhanced in Mg than high-$\alpha$ stars without planets. Summarizing the results obtained for the  Neptunian-like planet hosts, we can assume that
to form low-mass planets some elements, like Mg, might be particularly important. 

%
\begin{acknowledgements}

{This work was supported by the European Research Council/European Community under the FP7 through Starting Grant agreement number 239953. 
N.C.S. also acknowledges the support from Funda\c{c}\~ao para a Ci\^encia e a Tecnologia (FCT) through program Ci\^encia\,2007 funded by 
FCT/MCTES (Portugal) and POPH/FSE (EC), and in the form of grant reference PTDC/CTE-AST/098528/2008. 
V.Zh.A., S.G.S. and E.D.M are supported by grants SFRH/BPD/70574/2010, SFRH/BPD/47611/2008 and SFRH/BPD/76606/2011 from FCT (Portugal), respectively.
J.I.G.H., and G.I. acknowledge financial support from the Spanish Ministry project MICINN AYA2011-29060 and J.I.G.H.
also from the Spanish Ministry of Science and Innovation (MICINN) under the 2009 Juan de la Cierva Programme.}
\end{acknowledgements}

\end{document}